\documentclass[10pt,a4paper,twocolumn,showkeys,showpacs,pre,final]{revtex4-1}
\usepackage{dcolumn}
\usepackage[english]{babel}
\usepackage[utf8]{inputenc}     
\usepackage{amsmath,amssymb,amsfonts}
\usepackage{amsthm}
\usepackage{textcomp} 
\usepackage{xcolor}
\usepackage{colortbl}
\usepackage{comment} 
\usepackage{hyperref}
\hypersetup{
   pdfborder={0 0 0},            
   pdftoolbar=true,              
   pdfmenubar=true,              
   pdffitwindow=false,           
   pdfstartview={FitH},          
   pdfauthor={Lucas N. Jorge}, 
   pdftitle={Modelo de artigo},  
   pdfsubject={Modelo},          
   pdfkeywords={Modelo,artigo},  
   pdfproducer={LaTeX},    
   pdfcreator={pdfLaTeX},  
   pdfnewwindow=true,      
   colorlinks=true,       
   linkcolor=red,          
   citecolor=green,        
   filecolor=magenta,      
   urlcolor=cyan           
}
\usepackage{graphicx} 
\DeclareGraphicsExtensions{.eps,.png,.jpg,.pdf}
\graphicspath{
{Figs/} 
}
\usepackage{fancyhdr}

\definecolor{gold}{rgb}{0.85,.66, 0}
\definecolor{azul_escuro}{rgb}{0.12,0.26, 0.36}
\definecolor{azul}{rgb}{0.21,0.55, 0.0}
\definecolor{Azul}{rgb}{0.21,0.85,0.85}
\definecolor{laranja}{rgb}{0.95,.65, 0.20}
\definecolor{amarelo}{rgb}{0.95,.95, 0.55}
\definecolor{amarelo1}{rgb}{1,0.97, 0.20}
\definecolor{amarelo2}{rgb}{0.85,0.55, 0.0}
\definecolor{amarelo3}{rgb}{1.0,1.0, 0.80}
\definecolor{Amarelo}{rgb}{0.99,0.99, 0.66}
\definecolor{verde}{rgb}{0.57,0.94, 0.80}
\definecolor{Cinza50}{gray}{0.50}
\definecolor{Cinza10}{gray}{0.10}
\definecolor{Cinza}{gray}{0.80}
\definecolor{cinza}{rgb}{0.9,0.9,0.9}
\newlength{\LC}
\newcolumntype{A}{>{\columncolor[gray]{0.96}\raggedright\color{blue}}l}
\newcolumntype{B}{>{\columncolor[gray]{0.96}\raggedright\color{blue}}r}
\newcolumntype{F}{>{\columncolor[gray]{0.95}\raggedright}m{\LC}}
\newcolumntype{G}{>{\columncolor[gray]{0.9}\raggedright}m{1.70cm}}
\newcolumntype{K}{>{\columncolor[gray]{0.8}\raggedright}c}
\newcolumntype{H}{>{\columncolor{yellow}\raggedright}c}

\begin{document}

\preprint{APS/123-QED}

\title{The three-dimensional Baxter-Wu Model}
\author{L. N. Jorge}
\altaffiliation{Instituto Federal do Mato Grosso - Campus Cáceres, CEP. 78200-000, Cáceres, Mato Grosso. Brazil}
\email{lucasnjorge@gmail.com}
\author{L. S. Ferreira}
\altaffiliation{Instituto de Física, Universidade Federal de Goiás, C.P. 131 CEP 74001-970, Goiânia, Goiás, Brazil.}
\email{lucas.if.ufg@gmail.com}
\author{A. A. Caparica}
\altaffiliation{Instituto de Física, Universidade Federal de Goiás, C.P. 131 CEP 74001-970, Goiânia, Goiás, Brazil.}
\email{caparica@ufg.br}
\begin{abstract}
A classic three-dimensional spin model, based upon the Baxter-Wu scheme, is presented. It is found, by entropic sampling simulations,
that the behavior of the energy and magnetization fourth-order cumulants points out to a first order phase transition. 
A finite-size procedure was performed, confirming that the system scales with the dimensionality $d=3$, and yielding a high-resolution 
estimate of the critical temperature as $T_c=11.377577(39)$.
\end{abstract}

\pacs{Valid PACS appear here}
\maketitle

The study of three-dimensional spin models in statistical physics have great importance in science of materials, since it can 
describe, predict, or even design real systems. We can cite the well-know Ising model that is used to describe the Fe-Al 
magnetic alloy\cite{Dias2009,Plascak2000a,Freitas2012alloy}, once it is arranged in a bcc structure being composed by two 
interpenetrating cubic lattices. For the disordered case, the phase-diagram has been described using the site-diluted spin-$1$ 
Blume–Capel model (BC) in a simple cubic arrangement via a mean-field renormalization group approach in the pair 
approximation\cite{Dias2011}. Such technique and model were also used to characterize Fe-Ni-Mn and Fe-Al-Mn alloys
\cite{PenaLara2009}. The Ising model is also used to construct metamagnets in thin film geometry models. These systems were 
studied via Wang-Landau procedure and by importance sampling Monte Carlo (MC) simulations in investigations of their equilibrium 
phase diagram\cite{Chou2011metamagnet}. Such model was extended to the understanding of non-equilibrium relaxation processes 
in Co/Cr superlattices\cite{Mukherjee2010probing}. Metamagnets compounds FeCl$_{2}$ and FeBr$_{2}$ have its properties 
simulated by a study of a three-dimensional spatially anisotropic Ising superantiferromagnet in the presence of a magnetic 
field\cite{Salmon2013monte}, where a rich phase diagram was constructed. There are in nature hexagonal arrangements, as for example, 
some magnetic systems. Based on MC simulations, Ma \textit{et al}, have presented results of a film model that is described by a 
three-dimensional layered honeycomb lattice\cite{Ma2011}. This kind of lattice was used by Wang \textit{et al} in the 
characterization of molecular based magnetic film AFe$^{II}$Fe$^{III}$(C$_2$O$_4$)$_3$\cite{Wang2012monte}. The most of the 
hexagonal magnetic materials are described via Ising model in a triangular lattice, as is the case of Ca$_3$CO$_2$O$_6$, where 
the steplike magnetization behavior is strongly dependent of the external field and temperature
\cite{Maignan2004steplike,Maignan2000steplike,Kudasov2006steplike}. Hexagonal nanoparticles and nanowires with a core-shell 
structure, like CuS/Cu$_2$S with mixed spin ($1/2;1$), and spin-$1$ Zn/Se, has been successfully predicted and synthesized
\cite{Rossler2004synthetic,Wang2017monte,Lv2017monte,Kaneyoshi2011nanowire,Masrour2015nanowire,Ahmed2009microemulsion,Chen2008water}.

Although there are so many models in hexagonal arrangements in nature or predicted in literature, a three-dimensional model via 
Baxter-Wu interactions is lacking. Therefore the present work aims modeling a three-dimensional system that obeys a three-spin 
interaction like in the two-dimensional Baxter-Wu model, and investigating the order of the phase transition and estimating the
subsequent thermodynamic properties, using entropic sampling simulations.  

Proposed by Wood and Griffiths\cite{Wood1972} in 1972 and exactly solved by Baxter and Wu
\cite{Baxter1973,baxter1974ising2,baxter1974ising}, the Baxter-Wu model is a spin model that considers terms of triple coupling 
between the spins. It consists in a magnetic system defined on a two-dimensional triangular lattice, where, for the spin-$1/2$ 
case, the spins variables can assume the values $\sigma=\pm 1$ and are located at the vertices of the triangles. The three spin 
interaction is governed by the Hamiltonian

\begin{equation}
 H_{BW}=-J\sum_{<i,j,k>}s_is_js_k, \label{eq.baxterwu}
\end{equation}
where $J$ is the nearest-neighbor coupling parameter between the spins that fixes the energy scale, 
and the sum extends over all triangular faces of the lattice. 

\begin{figure*}[t]
  \includegraphics[scale=0.9]{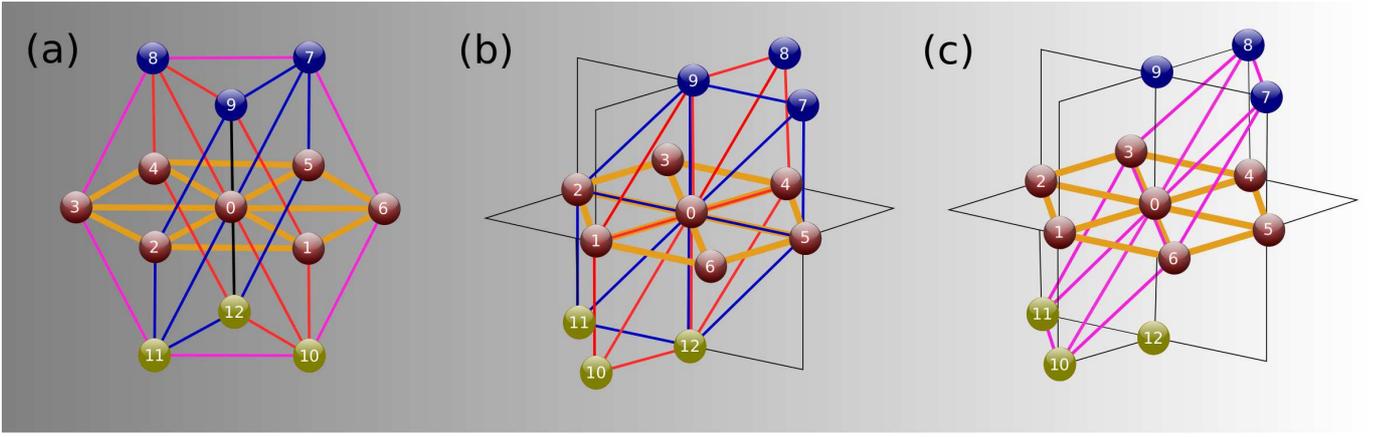}
\caption{(a) Three-dimensional lattice of the Baxter-Wu model. (b) Three-dimensional lattice of the Baxter-Wu model transposed 
into a cubic lattice. (c) The oblique plane.}
\label{fig1}
\end{figure*}

 To construct the three-dimensional version of the Baxter-Wu model, we consider a regular hexagon on the horizontal plane, with 
six spins in the vertices. In its center, there are three axes crossing it, in such a way that each one can be 
associated to a hexagon forming an angle of $60^{\oldstylenums{0}}$ with the initial plane, as shown in Fig. \ref{fig1}.(a).
In this figure we see that the initial hexagon is formed by the dots $(1,2,3,4,5,6)$ and in its center is the zero site.
The other three hexagons are formed by $(1,9,8,4,12,10)$, $(2,9,7,5,12,11)$ and $(3,8,7,6,10,11)$. So, a spin, that in the 
two-dimensional case has six nearest neighbors, and is surrounded by six triangular faces, has in this case $12$ nearest 
neighbors and counts with $24$ triangular faces surrounding it.

In this scheme, the three sites that belong to the upper plane, are located above the center of three alternated triangular 
faces, while the other three in the lower plane are located bellow the centers of the other three triangular faces, 
as shown in Fig. \ref{fig1}.(a). This lattice is known as hexagonal close-packed (hcp)\cite{Ashcroft}.

Therefore, the three-dimensional Baxter-Wu model is defined in a three-dimensional lattice with triangular 
interactions, with the energy given by

\begin{equation}\label{h3d}
 H_{BW3D} = -J\sum_{<i,j,k>}s_{i}s_{j}s_{k}, 
\end{equation}
where the sum extends over all possible triangular faces of the lattice and the spins variables are located 
at the vertices of the triangles and can assume the integer values $\sigma=\pm 1$. $J$ is the constant that scales 
the energy of the lattice, being the same in all directions. Unlike the two-dimensional version of the model, which
displays four ground state configurations -- one ferromagnetic and three ferrimagnetic -- the  
3D Baxter-Wu model has a single ground state configuration, namely the ferromagnetic one. When we try to construct a 
three-dimensional ferrimagnetic configuration, the triangular faces are satisfied for two planes, but for the third 
and fourth planes, frustrations appear, showing that it is impossible to obtain such ferrimagnetic constructions in 
three dimensions.

The Hamiltonian may then be decomposed in sums over four planes
\begin{align}
 H = &-J\left[\sum_{<i,j,k>}s_{i}s_{j}s_{k}\right]_{XY} -J \left[\sum_{<i,j,k>}s_{i}s_{j}s_{k}\right]_{XZ} \notag \\ 
 &-J\left[\sum_{<i,j,k>}s_{i}s_{j}s_{k}\right]_{YZ} -J\left[\sum_{<i,j,k>}s_{i}s_{j}s_{k}\right]_{Obl.},
\end{align}
which extend over all triangles of the lattice in each plane $XY$, $XZ$, $YZ$ (Fig. \ref{fig1}.(b)) and the 
oblique plane (Fig. \ref{fig1}.(c)), respectively.

In the cubic scheme, the energy of a particular configuration is given by
\begin{widetext}
$E = \frac{J}{3}[ \sum_{i=1}^{L}\sum_{j=1}^{L}\sum_{k=1}^{L}s_{i,j,k}(s_{i+1,j,k}s_{i,j-1,k}+s_{i,j-1,k}s_{i-1,j-1,k}
     +s_{i-1,j-1,k}s_{i-1,j,k}+s_{i-1,j,k}s_{i,j+1,k}
     +s_{i,j+1,k}s_{i+1,j+1,k}+s_{i+1,j+1,k}s_{i+1,j,k})$\\

     $ +\sum_{i=1}^{L}\sum_{j=1}^{L}\sum_{k=1}^{L}s_{i,j,k}(s_{i+1,j,k}s_{i,j,k+1}+s_{i,j,k+1}s_{i-1,j,k+1}
     +s_{i-1,j,k+1}s_{i-1,j,k}+s_{i-1,j,k}s_{i,j,k-1}
     +s_{i,j,k-1}s_{i+1,j,k-1}+s_{i+1,j,k-1}s_{i+1,j,k})$ \\

     $+\sum_{i=1}^{L}\sum_{j=1}^{L}\sum_{k=1}^{L}s_{i,j,k}(s_{i,j,k+1}s_{i,j-1,k}+s_{i,j-1,k}s_{i,j-1,k-1}
     +s_{i,j-1,k-1}s_{i,j,k-1}+s_{i,j,k-1}s_{i,j+1,k}
     +s_{i,j+1,k}s_{i,j+1,k+1}+s_{i,j+1,k+1}s_{i,j,k+1})$ \\
     
     $+\sum_{i=1}^{L}\sum_{j=1}^{L}\sum_{k=1}^{L}s_{i,j,k}(s_{i-1,j-1,k}s_{i-1,j,k+1}
     +s_{i-1,j,k+1}s_{i,j+1,k+1}+s_{i,j+1,k+1}s_{i+1,j+1,k}
     +s_{i+1,j+1,k}s_{i+1,j,k-1}+s_{i+1,j,k-1}s_{i,j-1,k-1}
     +s_{i,j-1,k-1}s_{i-1,j-1,k})]$,
 \end{widetext} 

where the division by three is because in this sums each triangular face is counted three times. 

In this work we adopted the order parameter as the total magnetization of the system, 
$M=\sum_{i,j,k=1}^{L}s_{i,j,k}$\cite{Jorge2016,Wagner2001,velonakis2013}, thus, in the simulations we picked only non-multiple 
of three lattice sizes.

\begin{figure*}[!t] 
\includegraphics[width=0.3\linewidth,angle=-90]{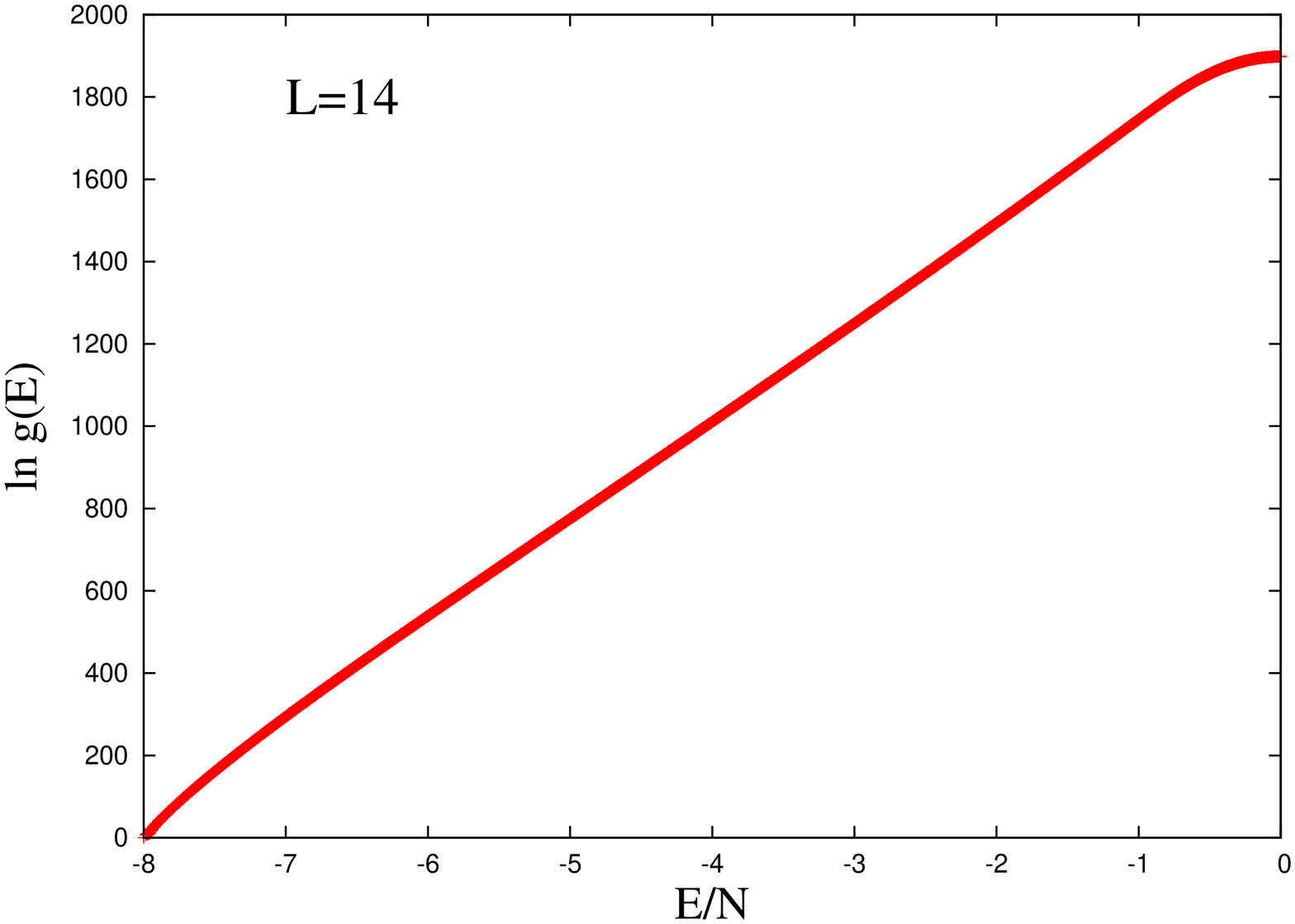}\hspace{0.5cm}
\includegraphics[width=0.3\linewidth,angle=-90]{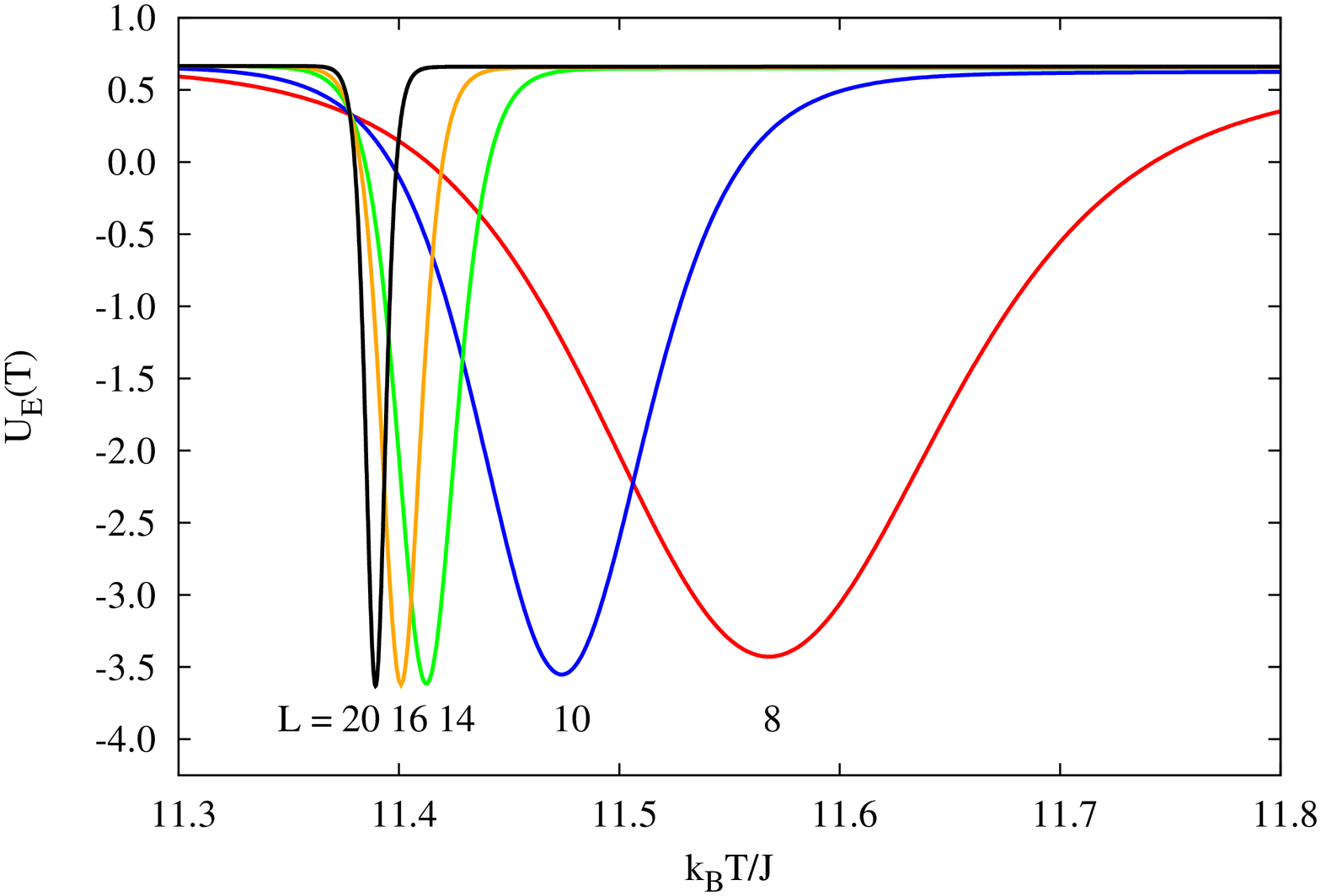}
\includegraphics[width=0.3\linewidth,angle=-90]{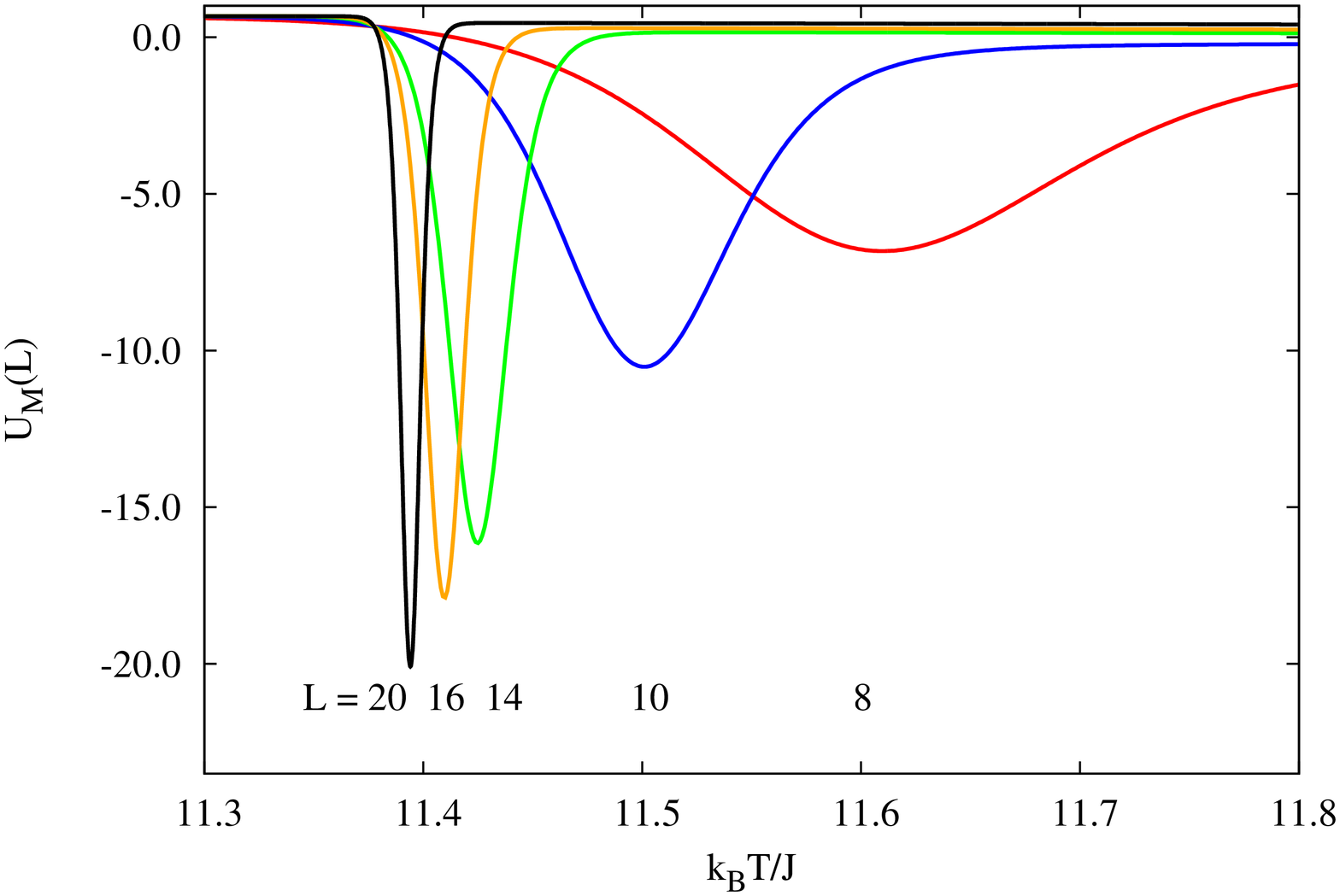}\hspace{0.5cm}
\includegraphics[width=0.3\linewidth,angle=-90]{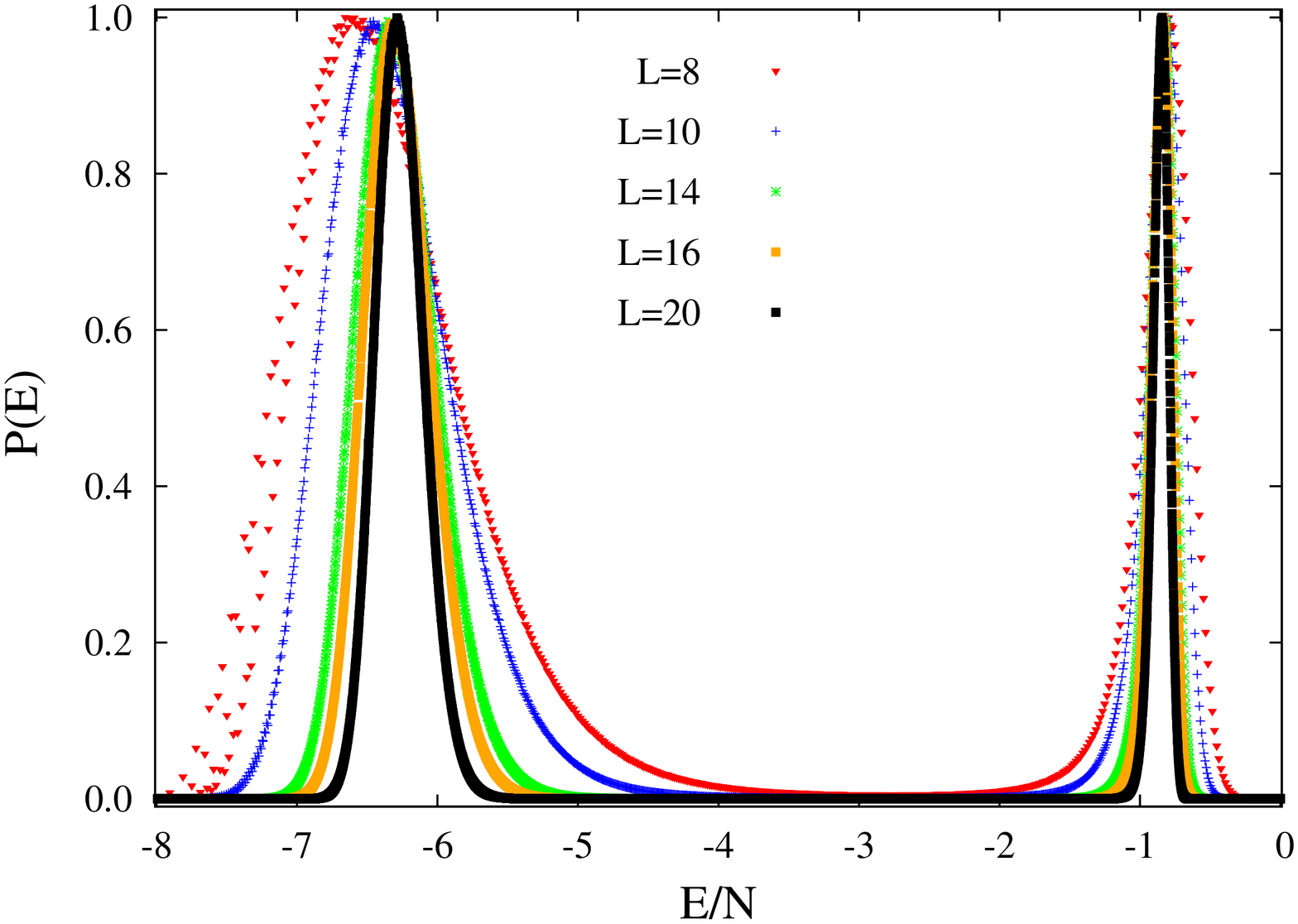}
\caption{(Top left) Logarithm of the density of states for the $L=14$. (Top right) Fourth-order energy cumulants as funtions of temperature.
(Bottom left) Fourth-order magnetization cumulants as functions of temperature. (Bottom right) Energy probability distributions as
function of the energy per particle.
}\label{fig2}
\end{figure*}

The entropic simulations applied to our model are based on the Wang-Landau method\cite{Wang2001}, that by means of a 
random walk in energy space allows the construction of the density of states $g(E)$, generating a flat histogram for the
energy distribution, and then the estimation of the canonical averages of any  thermodynamic quantities. In our simulations 
we include some improvements that enhance the accuracy and lead to substantial savings in CPU time. Namely, (i) we adopt 
the Monte Carlo sweep before updating the density of states, avoiding taking into account highly correlated configurations, 
(ii) we begin to accumulate the microcanonical averages only from the eighth Wang-Landau level ($f_7$), such that we discard 
the initial configurations that do not match with those of maximum
entropy\cite{Caparica2012}, (iii) we use a checking parameter $\varepsilon$ for halting the simulation\cite{Caparica2014} 
(the computational process is halted if the integral of the specific heat over a range of temperature calculated with the 
current density of states during the simulations varies less then $10^{-4}$ during a whole Wang-Landau level), and (iv) 
we begin all simulations, for all lattice sizes, beginning from the outputs of a single run up to the Wang-Landau level 
$f_6$, because up to this point the current density of states is not biased yet and can proceed to any final result that 
would be obtained beginning from the first Wang-Landau level $f_0$\cite{Ferreira2018}, a procedure that allows saving 
about $60\%$ of CPU time.

We carried out entropic simulations for $L\times L\times L$ lattices, picking $L=8,~10,~14,~16$ and $20$, with 
$n=24,~20,~20,~16$ and $16$ independent runs, respectively. We performed independent simulations on five such sets, so
that all our final results and error bars are taken as averages over the results of these sets. 

In Fig. \ref{fig2}.(a) we show the logarithm of the density of states of the lattice size $L=14$.

\begin{figure*}[!t] 
\includegraphics[width=0.3\linewidth,angle=-90]{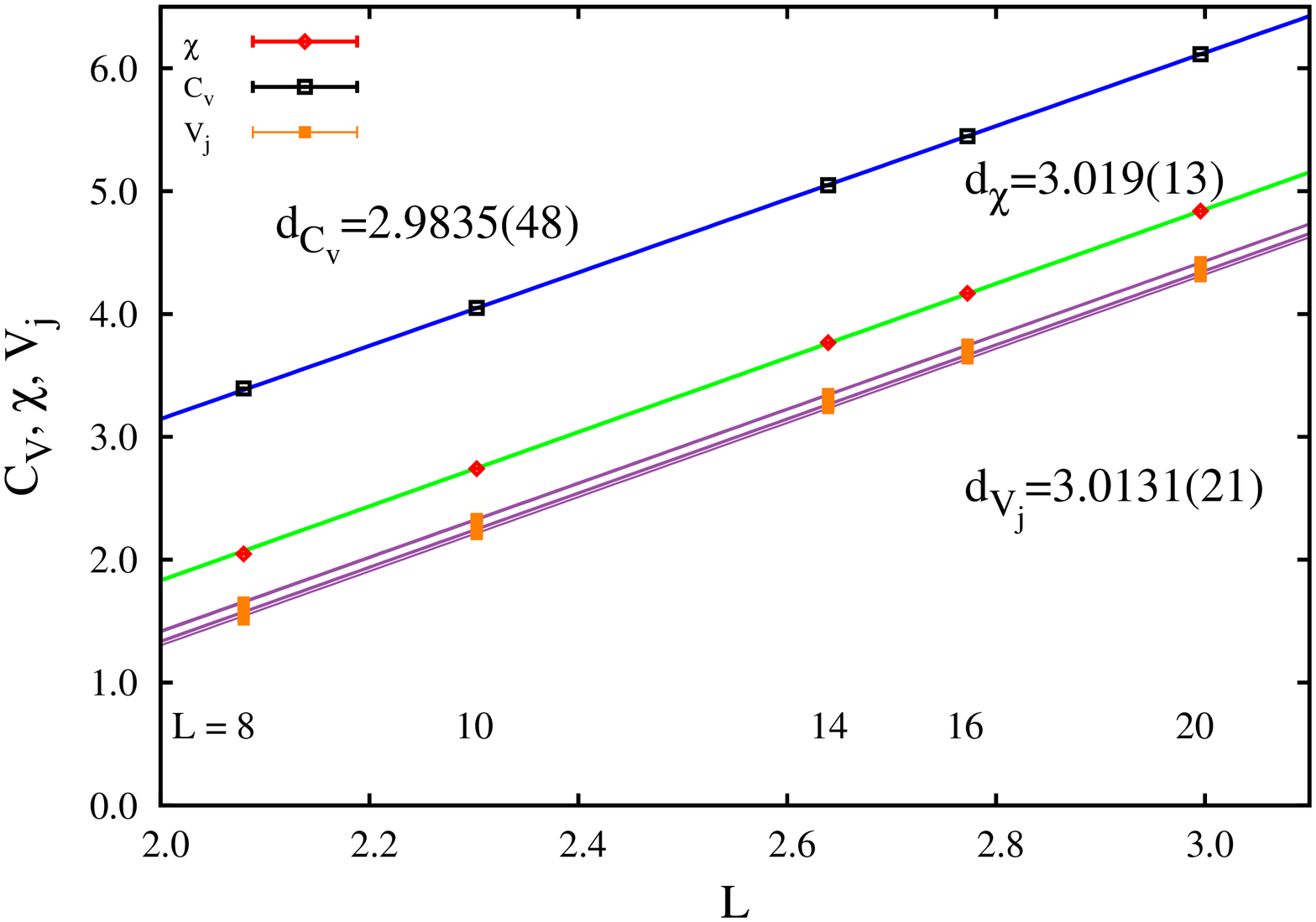}\hspace{0.5cm}
\includegraphics[width=0.3\linewidth,angle=-90]{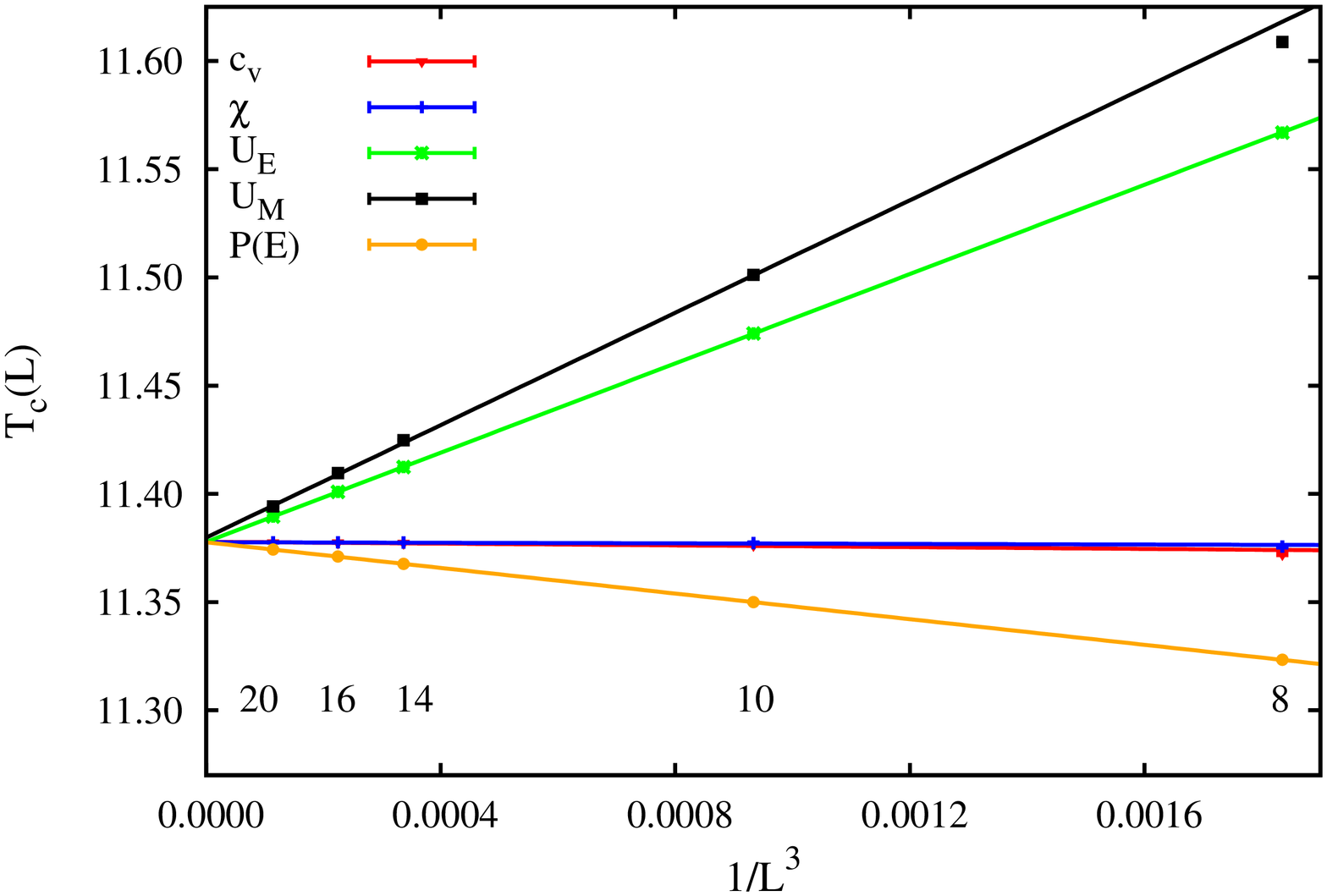}
\caption{(Left) Log-log plot of the maxima of the specific heat, the susceptibility, and the cumulants $V_j$ with the lattice size. 
The linear coefficients are close to the dimensionality of the system. (Right) Dependence on temperature of the minima of the 
energy and magnetization cumulants, the maxima of the specific heat and the susceptibility, and the double peaks of same high 
of energy density of probability, against $1/L^3$.} \label{fig3}
\end{figure*}

The behavior of the fourth-order energy and magnetization cumulants
\begin{equation}
 U_X(L)=1-\frac{\langle X^4\rangle}{3\langle X^2\rangle^2}, ~~X\equiv E, M. 
\end{equation}
shown in Figs. \ref{fig2}.(b) and \ref{fig2}.(c) give us solid evidences that our novel model undergoes a 1st order phase
transition. The energy cumulants intersect at a point close to the transition temperature, while the magnetization cumulants
exhibit sharp inverted minima, as expected in a discontinuous phase transition. In addition, the energy probability distributions
display double peaks of same high at the finite-size transition temperature, with a null probability valley between them, 
as we see in Fig. \ref{fig2}.(d).

In a system that suffers a discontinuous phase transition it is expected that the maxima of the specific heat and 
the magnetization should scale with the dimensionality. Another quantity that displays the dimensionality of the 
system in a discontinuous phase transition is $1/\nu$ in \cite{Ferrenberg1991,Caparica2000a,Caparica2015d,Jorge2016}

\begin{equation}
 V_j\approx \frac{1}{\nu}\ln L + \mathcal{V}_j(tL^{\frac{1}{\nu}}).\label{eq:Vj}
\end{equation}

In Fig. \ref{fig3} (left) we present this finite-size scaling 
behavior for the first of the five sets simulated. The values of all samples are shown 
in the first three columns of Table \ref{tab:3d}, where the final averages are given at the last line.
As usual in these entropic sampling procedures we neglect the error bars and calculate the averages
using the central values\cite{Caparica2014,Jorge2016,Ferreira2018}. 
\begin{table}[!h]       
   \setlength{\arrayrulewidth}{2\arrayrulewidth}  
\begin{tabular}{cccc}
\hline
\hline
    $d_{C_v}$~~  &   $d_{\chi}$~~ & $d_{V_j}$ &~~    $T_c$           \\ \hline
 2.9835(48)~~~~  &  3.019(13)~~~~ & 3.0131(21)~~~~ &  11.377618(30)   \\
 2.9813(48)~~~~  &  3.020(11)~~~~ & 3.0140(22)~~~~ &  11.377486(41)   \\
 2.9810(45)~~~~  &  3.018(11)~~~~ & 3.0141(22)~~~~ &  11.377690(17)   \\
 2.9815(44)~~~~  &  3.024(12)~~~~ & 3.0154(23)~~~~ &  11.377600(34)   \\
 2.9829(40)~~~~  &  3.017(11)~~~~ & 3.0135(20)~~~~ &  11.377489(36)   \\ \hline
2.98205(50)~~~~  & 3.0197(12)~~~~ & 3.0140(10)~~~~ &  11.377577(39)   \\  
\hline
\hline
\end{tabular}
\caption{Five finite-size scaling results yielding exponents close to the dimensionality for the maxima of
specific heat, the susceptibility, and the cumulants $V_j$, and for the critical temperature. The averages over 
all runs are displayed at the last line.}\label{tab:3d} 
\end{table}
In order to obtain our final results for $d$, the dimensionality, we again neglect the error bars and take an 
average of the three final results, yielding $d_{C_v, \chi, V_j }=3.005(12)$.

According to Fisher and Berker \cite{Fisher1982}, in first order transitions all finite size scaling procedures are made
in terms of powers of the lattice size, $L^{-d}$. Once confirmed that the system scales with the dimensionality, we can 
proceed with the determination of the critical temperature as the extrapolation for $L\rightarrow\infty$ ($L^{-d}=0$) of 
the best linear fits of the temperatures of the maxima of the specific heat and the susceptibility, the minima of the 
energy and magnetization fourth-order cumulants, and the temperatures where the energy probability distribution displays 
double peaks of same high. In Fig. \ref{fig3} (right) we depict these best fits for the first set of simulations. 
The mean critical temperatures for each set are displayed in Tab. \ref{tab:3d}, with the best estimate appearing in the 
last line, yielding $T_c=11.377577(39)$. This new three-dimensional model may consist of an useful platform for the simulation of 
existing compounds in nature.

\begin{acknowledgements}

We acknowledge the computer resources provided by LCC-UFG and IF-UFMT. L. N. 
Jorge and L. S. Ferreira acknowledge the support by FAPEG and CAPES, respectively.
\end{acknowledgements}
%
\end{document}